# Ytterbium- and chromium-doped fibre laser: from chaotic self-pulsing to passive Q-switching.

<u>Bernard Dussardier</u>[1], Jérôme Maria[1], Pavel Peterka[2]
[1]LPMC, Univ. de Nice-Sophia-Antipolis, UMR 6622, Parc Valrose, 06108 Nice Cedex 2, France
[2]IPE, Acad. of Science of the Czech Republic, v.v.i., 18251 Prague, Czech Republic
*bernard.dussardier@unice.fr      http://lpmc.unice.fr/*

**Abstract:** A spontaneously chaotic, self-pulsing ytterbium-doped fibre laser is partially stabilized into the passively Q-switched mode of operation using a chromium-doped saturable absorber fibre. This original all-fibre laser produces sustained and stable trains of smooth pulses at high repetition rate.

**Introduction**: Q-switched fibre lasers are alternatives to bulk lasers operating in the ns- to µs-range for many applications such as material processing, medicine, remote sensing. However they use externaly driven bulk modulators or passive components in free space sections,causing alignment and reliablity problems. All-fibre self-pulsed laser are prefered, because they are versatile, reliable, factory-aligned and integrated. Some double-clad ytterbium-doped fibre lasers (DCYFL) produce random self-pulsing along two main regimes : the sustained selfpulsing (SSP) and self mode locking (SML) [i]. Developing "all-fibre" passively Q-switched fibre lasers (PQSFL) is interesting with transition metal (TM) ions-doped fibres used as saturable absorbers (SA). TM such as $Cr^{4+}$ offer strong, broad and fast relaxing absorption bands. We had proposed a core-pumped PQSFL [ii] using a Cr-doped fibre [iii] as SA (CrSAF). We take advantage of the self-pulsing in DCYFL and a CrSAF as stabilizer, resulting in an "all fibre" PQSFL based on $Yb^{3+}$ and $Cr^{4+}$ ions.

**Experiments and results** : The laser setup is described on Fig. 1. Without the CrSAF, the laser is chaotic over the pumping range (up to 16 W on the cleaved fiber) (fig.2, left). The pulse envelops are strongly modulated by short pulses at a cavity round-trip frequency. Drastic changes appear when inserting the CrSAF, including stable passive Q-switching (peak RMS fluctuations <10%) upto 16 W of pump (fig.2 right). The Q-switched envelopes are smoothed compared to the DCYFL without CrSAF. Typical characteristics are: 400 ns width, 350 kHz repetition, 2.6 µJ/pulse and 6.4 W peak power.

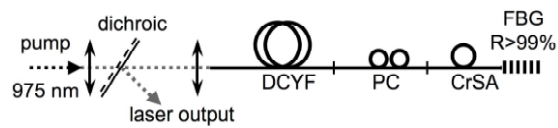

*Fig.1: Laser setup. DCYF: double-clad $Yb^{3+}$-doped fiber, PC: polarization controler, CrSA: Cr-doped saturable absorber fiber, FBG: highly reflecting fiber Bragg grating at 1064 nm, FWHM= 0.3 nm. Pump: 20-W multimode fiberized laser diode.*

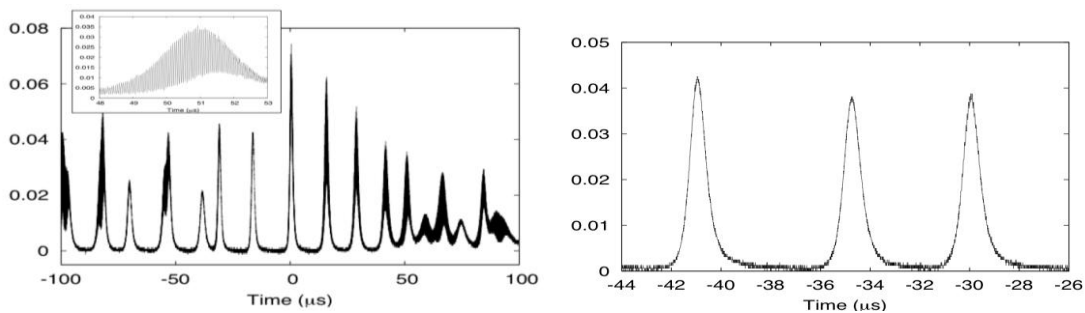

*Fig.2: Left: chaotic DCYDFL output. Right: stablilized PQSFL output*

**Discussion** : The laser output is typical of PQSFL, although the set of cavity parameters were causing chaotic behaviour without the CrSAF. The capacity of the CrSAF to stabilise SSP and cancel out SML in an DCYF laser has never been reported, to the best of our knowledge, and needs further studies including comprehensive, spatially and temporally resolved modelling. The device can be optimized to further saturate the CrSAF and reduce pulse duration. It could be further integrated and power upgraded (MOPA configuration) to permit application.

**Acknowledgements** : "Fédération Doeblin" (CNRS); grant MSMT-Kontakt ME10119 (Czech Rep.) (Czech Rep.); EGIDE 'Barrande' 17360VA. LPMC is with GIS 'GRIFON' (CNRS, http://grifon.xlim.fr/).

[i] F. Brunet et al., "A Simple Model Describing Both Self-Mode Locking and Sustained Self-Pulsing in Ytterbium-Doped Ring Fiber Lasers", *J. Light. Technol.* 23 2131 (2005)

[ii] L. Tordella et al., "High repetition rate passively Q-switched $Nd^{3+}$:$Cr^{4+}$ all-fibre laser", *Electron. Lett.*, **39**, 1307 (2003)

[iii] V. Felice et al., "Chromium-doped silica optical fibres : influence of the core composition on the Cr oxidation states and crystal field", *Opt. Mat.*, **16** 269 (2001)